\documentclass[aps,twocolumn,preprintnumbers,nofootinbib,superscriptaddress]{revtex4}
\usepackage{latexsym}
\usepackage{amsmath,amsfonts}
\usepackage{amsbsy}
\usepackage{mathrsfs}
\usepackage{color}
\usepackage{float}
\usepackage{dsfont}
\usepackage{psfrag}
\usepackage{enumerate}
\usepackage{soul}
\usepackage{gensymb}
\usepackage{bm}

\usepackage{hyperref}% add hypertext capabilities
\usepackage{color}
\usepackage{ulem}

\usepackage{amsmath,amssymb,calc,amsfonts}
\usepackage{latexsym}
\usepackage{hyperref}
\usepackage{graphicx,calc,epsfig}

\begin{document}

\title{Black Hole Shadows in Verlinde’s Emergent Gravity}

\author{Kimet Jusufi}\email{kimet.jusufi@unite.edu.mk}
\affiliation{Physics Department, State University of Tetovo, Ilinden Street nn, 1200,
Tetovo, North Macedonia}

\author{Saurabh}\email{sbhkmr1999@gmail.com}
\affiliation{Department of Physics, Dyal Singh College, University of Delhi-110003}

% Abstract of the paper
\begin{abstract}
We study the effect of baryonic matter and apparent dark matter on black hole shadow in Verlinde’s Emergent Gravity. To do so, we consider different baryonic mass profiles and an optically-thin disk region described by a gas in a radial free fall around the black hole.  Assuming that most of the baryonic matter in the galaxy is located near the galactic center surrounding a supermassive black hole, we use two models of power law mass profile for the baryonic matter to study the effect of apparent dark matter on the shadow and the corresponding intensity. We find that the effect of the surrounding matter on the shadow size using observational values is small, however, it becomes significant when the surrounding baryonic matter increases. To this end, we show that the effect of simple power law function in the limit of constant baryonic mass in Verlinde’s theory implies an apparent dark matter effect which is similar to the standard gravity having an isothermal dark matter profile.  We also find the intensity of the electromagnetic flux radiation depending on the surrounding mass. 
\end{abstract}

% Select between one and six entries from the list of approved keywords.
% Don't make up new ones.
%\begin{keywords} {Emergent gravity, Black holes, Shadows}

\maketitle

\section{Introduction}
Today's astrophysical observations seem to suggest that the galaxies contain supermassive black holes (SMBHs) at their galactic centers. From general relativity, we know that black holes are characterized by event horizon at the boundary and a singularity at the center. The presence of event horizon means that you can enter them but never exit. The most compelling evidence that can be linked with supermassive black holes is at the center of our own galaxy. At the center of the Milky Way galaxy, there is a black hole with a mass four million times the mass of the sun. It has been shown that black holes can be completely determined by the parameters; black hole mass, angular momentum and electric charge. In realistic astrophysical situations, around a supermassive black hole, there is an accretion mass which is due to the fact that a black hole (BH) can capture the light received from nearby stars or accretion disks into bound orbits. Among other things, black holes are characterized by a photon sphere radius which consists of orbiting light rays. Furthermore, the light rays can be unstable/stable if the photon can fall/escape to infinity, respectively. The strongest evidence supporting the existence of black holes is the shadow images of the M87 galactic center black hole reported by the Event Horizon Telescope (EHT) (\cite{Akiyama:2019bqs,Akiyama:2019eap}) and the detection of gravitational waves by LIGO \cite{Abbott:2016blz}.
	
Based on the properties of black hole shadow, one can test general relativity and different alternative theories of gravity (\cite{Psaltis:2020lvx}). It is thus important to note that the shadow images can shed light on many astrophysical problems, such as the accretion matter around a black hole including the dark matter distribution in the galactic center. In this respect, not only the distortion in shadow images due to the BH mass/spin is important, but also the effect of surrounding matter can be significant. Furthermore, one can use different spherical accretion models to study the intensity of the electromagnetic radiation as seen by an observer at a far distance from the black hole. That being said, the shadow images and intensity of electromagnetic radiation are very important tools to test the existence of both black holes and other exotic objects such as wormholes or naked singularities (\cite{Falcke:1999pj,1997AA...326..419J,PhysRevD.87.107501,rajibul,Dey:2020haf,Gyulchev:2019tvk}).
	
	Recently, Verlinde proposed emergent gravitational theory \cite{Verlinde:2016toy} according to which, dark matter can be viewed as an emergent manifestation of gravity. In this theory, the gravitational potential $\Phi(r)$ caused by enclosed baryonic mass distribution exceeds that of general relativity on galactic and larger scales. Furthermore, Verlinde argued that due to the contribution of baryonic mass to the gravitational potential, there exists an extra gravitational effect due to a volume law contribution to the entropy that is associated with positive dark energy in our universe. In this theory, the additional gravitational force can thus be understood as follows: the baryonic mass distribution reduces the entropy content of the universe (meaning that the total entropy associated with dark energy is maximal in a universe without matter due to the fact that it would be non-locally distributed over the entire space available), as a consequence of this removal of entropy due to matter, there is an elastic response of the underlying microscopic system. The important result here is that this effect has observational consequences on large scale structures as an additional gravitational force and provides an alternative way of describing the dark matter as an apparent dark matter distribution. 

Verlinde's theory was recently tested using weak gravitational lensing \cite{Brouwer:2016dvq}, radial acceleration relation \cite{Lelli:2017sul}, with early type galaxies \cite{Tortora:2017uid}, and galaxy cluster scales \cite{Tamosiunas:2019ghq}. Normally it is assumed that dark matter is some form of an elementary particle in some models, and plays a key role in many astrophysical processes, but as of today, it continues to be an unsolved problem in physics. There have been extensive studies in recent years on this matter. In recent papers, the effect of dark matter on BH shadow has been investigated (\cite{Sbh2020,Konoplya:2019sns,2020EPJC...80..354J,Boshkayev:2020kle}). In this paper, we aim to investigate the effect of surrounding baryonic mass along with apparent dark matter on black hole shadow. While doing this, we shall use the  physical properties of electromagnetic radiation emitted from an optically-thin disk with accretion flow in free fall around a spherically symmetrical black hole.
	 
	The structure of our paper is laid out as follows: In Section II we point out two models to describe the baryonic and apparent dark matter surrounding the black hole. In Section III and Section IV, we study the photon sphere and the shadow images along with the intensity of the radiation produced by a spherically thin medium described by an infalling gas model. In Section V, we use the EHT results to constrain the surrounding baryonic mass for Model I and Model II, respectively. Finally, in Section VI, we comment on our results. 
%\begin{figure*}
%    \centering
%    \includegraphics[scale=0.6]{panelmet2.pdf}
%    \caption{First row : $M_b= 0.1, 0.5, 1.0$, Second row: $M_b=1.1, 1.5, 2$}
 %  \label{fig:my_label}
%\end{figure*}

\section{Black hole surrounded by matter}

\subsection{Model I}
In a recent paper, Eric Verlinde (\cite{Verlinde:2016toy}) has proposed a novel emergent gravitational theory.
The most important claim of the theory is that dark matter has no particle origin but instead is an emergent manifestation in modified gravity. Assuming spherical symmetry, Verlinde showed that
\begin{equation}
\int_0^r \frac{M^2_D(r') dr}{r'^2}=\frac{a_0 M_B(r) r }{6}
\end{equation}
with $a_0=c H_0$, where $H_0=2.36\times 10^{-18}~{\rm s^{-1}}\simeq \sqrt{\Lambda/3}$
is the current Hubble parameter, $\Lambda$ is the cosmological
constant, and  $M_B(r)$ ($M_D(r)$)
is the baryonic mass (dark mass) inside a sphere of
radius $r$. This equation describes the amount of apparent dark matter $M_D(r)$ in terms of the amount of baryonic matter $M_B(r)$ for spherically symmetrical case consisting of stars
of mass, ionized gas of mass and neutral hydrogen of mass. Eq. (2.1) can be also rewritten as follows
\begin{equation}
M_D^2(r)=\frac{a_0 r^2}{6}\frac{d}{dr}(r M_B(r)).
\end{equation}

%We shall assume here that the metric describing the geometry near a black hole which is given by
Now we proceed to solve the Tolman-Oppenheimer-Volkoff (TOV) equation in the dark matter halo with a black hole. To analyze the properties of the system composed of a black hole and dark matter envelope, the line element is chosen in the standard static and spherically symmetric form as
\begin{equation}\label{eq:le}
d s^2=-e^{\lambda(r)} d t^2 + e^{\nu(r)}d r^2 + r^2 \left(d \theta^2 + \sin^2 \theta d\phi^2\right),
\end{equation}
where $(t,r,\theta,\phi)$ are the time and spherical coordinates, $\lambda(r) $ and $\nu(r)$ are the sought metric functions. It is known that we can always chose
\begin{eqnarray}\notag
e^{-\nu(r)}&=& 1-\frac{2 M(r)}{r}.
\end{eqnarray}

To find the full solution we need to solve the TOV equations, which read
\begin{eqnarray}
\frac{d P(r)}{d r}&=&-(\rho(r) +P(r)) \frac{M(r)+4 \pi r^3 P(r)}{r(r-2 M(r))} ,\\
\frac{d \lambda(r)}{dr}&=& 2\,\frac{M(r)+4 \pi r^3 P(r)}{r(r-2 M(r))} ,
\end{eqnarray}
where $P(r)$ is the dark matter pressure.
At this point, it is important to note that we shall neglect the cosmological constant in Einstein field equations. We should, hence, obtain a Schwarzschild solution given by the above metric function when the surrounding matter is absent. The total mass function can be considered as the sum of black hole mass, baryonic mass, and apparent dark matter 
\begin{equation}
    M(r)=m+M_B(r)+M_D(r).
\end{equation}
In the subsequent section, we are going to consider different mass to obtain the spacetime geometry around the black hole. 

\begin{figure}
  \centering 
  \includegraphics[scale=0.8]{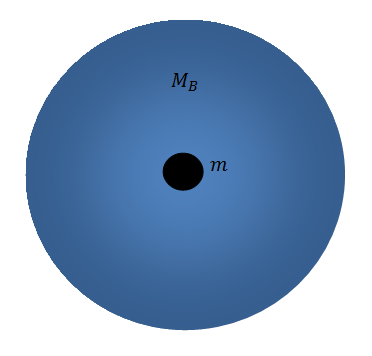}
    \caption{Schematic representation  of the galactic central  region having a black hole and baryonic matter with mass $M_B$.}
\end{figure}

\subsubsection{Case I }
Let us begin by considering a simple power law profile for the baryonic mass profile $M_{B}(r)$ given by
\begin{equation}
M_{B}(r)=M_B\left(\frac{r}{r+r_c}\right).
\end{equation}
As a special case when $r_c \to 0$ we obtain the point mass approximation $M_B(r)=const$. Using this equation along with Eq. (2.2) we find the corresponding apparent dark matter mass
\begin{equation}
M_{D}(r)=\frac{v_0^2 r^2}{r+r_c}\sqrt{1+\frac{2 r_c}{r}}=\frac{v_0^2 r^2}{r+r_c}\Big[1+\frac{r_c}{r}+\hdots \Big],
\end{equation} 
where we have defined $v_0^2=\sqrt{a_M M_B}$. In particular, if we use the above approximation for the apparent dark matter mass and we fix the constant such that the metric function vanishes at the horizon $r_h=2m$, we find
\begin{eqnarray}
e^{-\nu(r)}&=& 1-\frac{2 m}{r}-2 M_B\left(\frac{1}{r+r_c}-\frac{1}{2m+r_c}\right)\\\notag
&-& 2 v_0^2\Big[\frac{ r}{r+r_c}\left(1+\frac{r_c}{r}\right)-\frac{ 2m}{2m+r_c}\left(1+\frac{r_c}{2m}\right)\Big].
\end{eqnarray}

Using Eqs. (2.7)-(2.8) along with  Eq. (2.5) 
%\begin{equation}
%\lambda(r)=\frac{2(r_c v_0^2 -M_B)\ln|2 v_0^2 r+2M_B  -r-r_0|-4 \ln(r) (v_0^2-1/2) M_B}{  (2v_0^2-1)(2M_B-r_c)}.
%\end{equation}
when the black hole is introduced, we can simplify the metric by fixing the constant such that the metric vanishes at the horizon $r_h=2m$, yielding the following result
\begin{eqnarray}
    e^{\lambda}&=& 1-\frac{2m}{r}\\\notag
&+& \exp\Big[{\frac{-2M_B \zeta_1 -4 (v_0^2 r_c+M_B) \ln(r)(v_0^2-1/2)}{4 r_c v_0^4+4(M_B-r_c)v_0^2-2M_B+r_c}}\Big]\\\notag
    &-& \exp\Big[{\frac{-2M_B \zeta_2 -4 (v_0^2 r_c+M_B) \ln(r_h)(v_0^2-1/2)}{4 r_c v_0^4+4(M_B-r_c)v_0^2-2M_B+r_c}}\Big]
\end{eqnarray}
where 
\begin{eqnarray}
\zeta_1&=&\ln\left(|2(r+r_c)v_0^2+2M_B-r-r_c|\right),\\
\zeta_2&=&\ln\left(|2(r_h+r_c)v_0^2+2M_B-r_h-r_c|\right).
\end{eqnarray}
%\begin{equation}
%    e^{\lambda}=1-\frac{2m}{r}-\frac{2M_B}{r_c}\Big[\ln(\frac{r+r_c}{r})-\ln(\frac{2m+r_c}{2m})\Big]+2 v_0^2 \ln(\frac{r+r_c}{2m+r_c}).
%\end{equation}
This means that the distribution of matter is not very close to the black hole hence its horizon remains unaltered. Furthermore, we obtain the matching of the  $e^{\lambda(r)}=e^{-\nu(r)}$ at the horizon. We also note the sign difference before $v_0^2$ in the $e^{\lambda(r)}$ and $e^{-\nu(r)}$ which is a result of the relativistic TOV equation. In out setup we have  used the simplest case for dark matter having EoS with $\omega=0$.  It is easy to see that in the limit $r_c \to 0$, we recover the point mass approximation.   It was already shown in the context of Verlinde's theory (\cite{Liu:2016nwt}) that a collapsing matter will eventually form a black hole with a nontrivial global topology  similar to the global monopole metric originally found by \cite{Barriola:1989hx}. They have solved Einstein's field equations by considering  a  matter contribution, say, a star of constant density, which led to such nontrivial topology in large scale structures. It is also observed that the $e^{\lambda(r)}$ term diverges as $r \to \infty$, however, we note that a cut-off distance must be introduced at some distance $r=r_{\star}$, resulting with some finite effective mass $M_{eff}$ which gives the total mass enclosed by the system.

\begin{figure*}
    \includegraphics[scale=0.62]{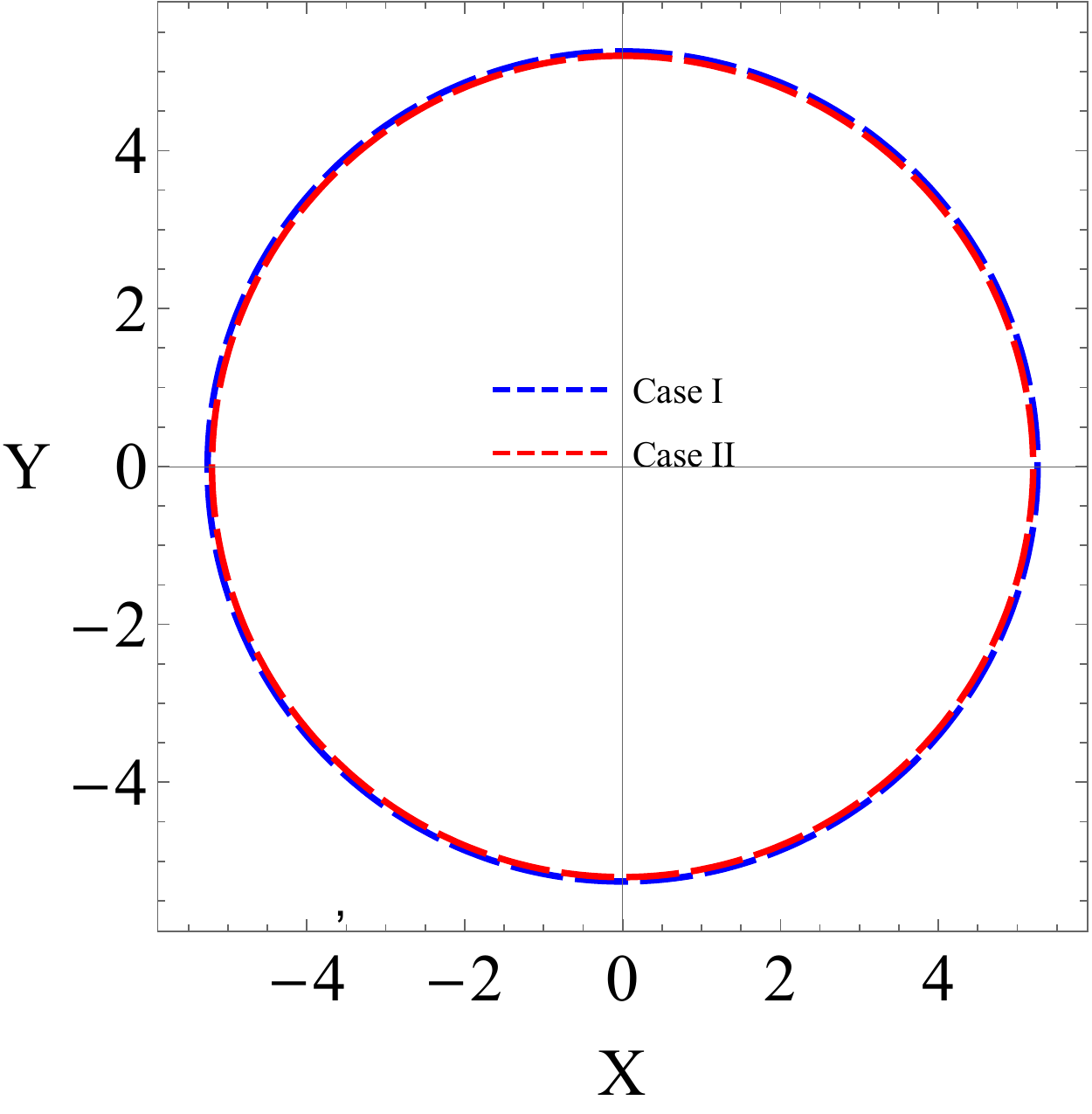}
        \includegraphics[scale=0.62]{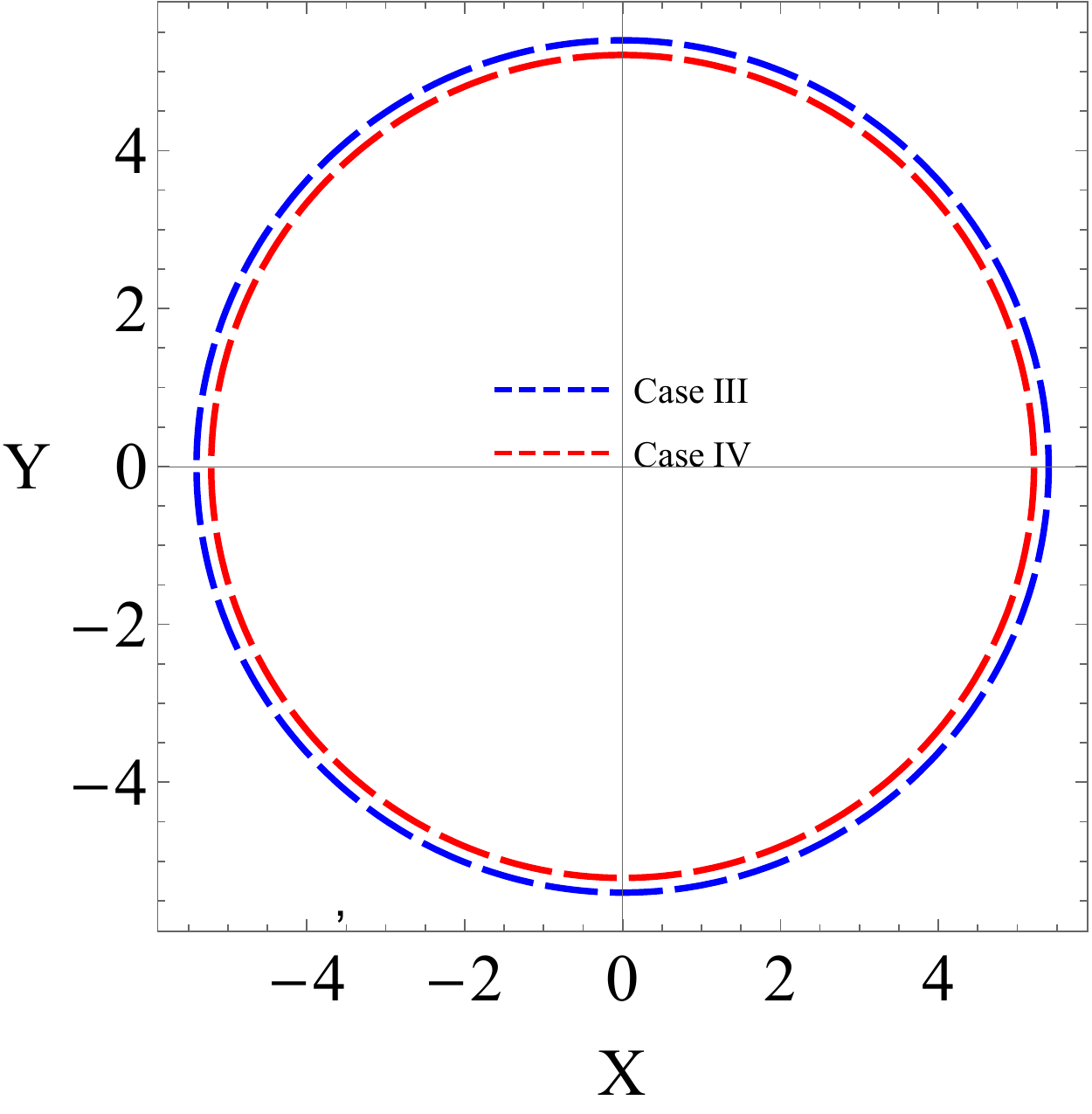}
    \caption{Left panel: Black hole shadows using Case I ans Case II models. Right panel:Black hole shadows using Case I ans Case II models.   In both cases we have used $r_c=10^6, M_B=10^3$ measured in units of the black hole mass and $v_0=0.001$. }
\end{figure*}

\subsubsection{Case II}
Let us proceed further by considering a power law profile for the distribution of baryonic matter around the black hole which is given by the so-called Hernquist model (see, \cite{Zhao:1995cp})
\begin{equation}
    M_B(r)=M_B \frac{r^2}{(r+r_c)^2}.
\end{equation}

Using Eq. (2.2) we find the apparent dark matter mass to be
\begin{equation}
    M_D(r)=\frac{v_0^2 r^{5/2}}{(r+r_c)^{3/2}} \sqrt{1+\frac{3 r_c}{r}}=\frac{v_0^2 r^{5/2}}{(r+r_c)^{3/2}} \left( 1+\frac{3 r_c}{2 r}+\hdots \right),
\end{equation}
where $r_c$ is the core radius of the baryonic matter. Using the last two equations we can approximate the solution as follows
\begin{equation}
e^{-\nu(r)}= 1-\frac{2 m}{r}-2 M_B\left(\frac{r}{(r+r_c)^2}-\frac{2m}{(2m+r_c)^2}\right)\notag
\end{equation}
\begin{equation}
-2 v_0^2\Big[\frac{ r^{3/2}}{(r+r_c)^{3/2}}\left(1+\frac{3r_c}{2r}\right)-\frac{ (2m)^{3/2}}{(2m+r_c)^{3/2}}\left(1+\frac{3r_c}{4m}\right)\Big].
\end{equation}
It can easily be seen that we have imposed the condition that $e^{-\nu(r)}$ vanishes at the event horizon $r_h=2m$ when the black hole is introduced. In a similar way, we can approximate the solution as follows
\begin{eqnarray}
    e^{\lambda}&=& 1-\frac{2m}{r} \\\notag
&+& e^{-\frac{2 M_B}{r+r_c}}  \exp{\Big[ \frac{2 v_0^2 \left(  \Delta_1 (r+r_c)+\sqrt{r(r+r_c)}\right)} {r+r_c}\Big]}\\\notag
    &-& e^{-\frac{2 M_B}{r_h+r_c}}  \exp{\Big[ \frac{2 v_0^2 \left(  \Delta_2 (r_h+r_c)+\sqrt{r_h(r_h+r_c)} \right)} {r_h+r_c}\Big]} 
\end{eqnarray}
%\begin{eqnarray}\notag
 %   e^{\lambda}&=& 1-\frac{2m}{r}-2M_B \left(\frac{1}{r+r_c}-\frac{1}{2m+r_c}\right)\\\notag
  %  &+& 2 v_0^2 \Big(\frac{y_1 (r+r_0)+\sqrt{r(r+r_0)} }{r+r_0}\Big) \\
 %    &-& 2 v_0^2 \Big(\frac{y_2 (2m+r_0)+\sqrt{2m (2m +r_0)} }{2m +r_0}\Big)
%\end{eqnarray}
where it has been defined
\begin{eqnarray}
\Delta_1&=&\ln\left(r_c/2+r+\sqrt{r (r +r_c)}\right),\\
\Delta_2&=&\ln\left(r_c/2+r_h+\sqrt{r_h (r_h +r_c)}\right).
\end{eqnarray}
Again, it can be observed that Eq. (2.16), by construction, vanishes at the event horizon.

\subsection{Model II}
In this model, we are going to use a different way to construct the spacetime metric around the black hole. Let us start by writing the total acceleration on a test particle which is given by
\begin{equation}
a=a_B+a_{D},
\end{equation}
where the effect of baryonic matter and aparent dark matter can be written as
\begin{equation}
a_B=\frac{M_B(r)}{r^2},\,\,\,\,\,a_{DM}=\frac{M_{D}(r)}{r^2}.
\end{equation}

One can now find tangential velocity of a test particle moving
in the dark halo in spherically symmetrical space-time using the well known relation
\begin{equation}
v^2_{tg}(r)=\frac{M_B(r)}{r}+\sqrt{a_M \frac{d}{dr}(r M_B(r))},
\end{equation}
where 
\begin{equation}
a_M=\frac{a_0}{6},
\end{equation}
and
\begin{equation}
a_0=5.4 \times 10^{-10} \text{m/s}^2.
\end{equation}

\subsubsection{Case III}
The simplest scenario is to consider the surrounding baryonic matter given by the power law (2.7). Here we are assuming that the matter is located entirely near the galactic center. For the tangential velocity, we find that
\begin{equation}
v^2_{tg}(r)=\frac{M_B }{r+r_c}+\frac{v_0^2 r}{r+r_c}\left(1+\frac{r_c}{r}\right),
\end{equation}

We can now consider a static and spherically symmetrical spacetime ansatz written in Schwarzschild coordinates as follows 
\begin{equation}
ds^{2}=-f(r)dt^{2}+\frac{dr^{2}}{g(r)}+r^{2}\left(
d\theta ^{2}+\sin ^{2}\theta d\phi ^{2}\right).
\end{equation}
Given the tangential velocity, one can calculate radial function $f(r)$ from the following equation
\begin{align}
      v_{tg}^{2}\left( r\right) =\frac{r}{\sqrt{f(r)}}\frac{d \sqrt{f(r)}}{dr}=r \frac{d\ln(\sqrt{f(r)})}{dr}.
\end{align}

Using the above equation, we obtain 
\begin{equation}
f(r)=C\,  r^{2 v_0^2}\,\left({1+\frac{r_c}{r}}\right)^{-\frac{2M}{r_c}}.
\end{equation}
where $C$ is a constant of integration. Before we introduce the black hole in our model, let us show an interesting result when $r_c \to 0$, that is the point mass approximation $M_B=const.$ corresponds to the same physical situation of having apparent dark matter described by the isothermal sphere of dark matter profile in general theory of relativity. To see this, we introduce the average mass densities $\bar{\rho}_B(r)$ and $\bar{\rho}_D(r)$
inside a sphere of radius $r$ by writing the integrated masses. The total mass density profile inside the core is
then given by
\begin{eqnarray}
M_B(r)&=&\frac{4 \pi}{3}r^3 \bar{\rho}_B(r),\\
M_D(r)&=&\frac{4 \pi}{3}r^3 \bar{\rho}_D(r)
\end{eqnarray}

Following Verlinde (\cite{Verlinde:2016toy}), we can introduce the slope parameters
\begin{eqnarray}
\bar{\beta}_B(r)&=&-\frac{d\log \bar{\rho}_B(r)}{d \log r}, \\
\bar{\beta}_D(r)&=&-\frac{d\log \bar{\rho}_D(r)}{d \log r}.
\end{eqnarray}
respectively. One finds that the average apparent dark matter
density obeys
\begin{equation}
\bar{\rho}_D(r)^2=(4-\bar{\beta}_B(r))\frac{a_0}{8 \pi r}\bar{\rho}_B(r)
\end{equation}

Finally, given the average mass density $\bar{\rho}_D(r)$ one can find the actual mass density $\rho_D(r)$
for apparent dark matter via the relation
\begin{equation}
\rho_D(r)=(1-\frac{1}{3}\bar{\beta}_D(r))\bar{\rho}_D(r).
\end{equation}

Let us now focus on a particular example by taking:  $\bar{\beta}_B=3, \bar{\beta}_D=2$. In this way, it follows that
\begin{equation}
\bar{\rho}_D=\frac{C}{r^2},
\end{equation}
where $C$ is a constant. Similarly, we have an additional equation
\begin{equation}
\rho_D=\frac{1}{3}\frac{C}{r^2}.
\end{equation}
Taking the constant $C=3v_0^2/4 \pi$, we obtain
\begin{equation}
\rho_D=\frac{v^2_0}{4 \pi r^2},
\end{equation}
that is the well known isothermal sphere of dark matter profile in normal gravitational theory with the total dark matter mass inside a sphere of radius $r$ which is given by
\begin{equation}
M_{D}(r)=4 \pi \int_0^r \rho_D(r') r'^2 dr'=v_0^2 r.
\end{equation}
Directly using equation (2.1), we find that the corresponding baryonic mass is equal to 
\begin{equation}
M_{B}=\frac{v_0^4}{a_M}.
\end{equation}
In other words, we end up with a constant baryonic mass. If we now make use the equation for tangential velocity, it follows that
\begin{equation}
v^2_{tg}(r)=\frac{v_0^4}{a_M r}+v_0^2.
\end{equation}
On solving the last equation we find, 
\begin{equation}
f(r)= \left(\frac{r}{r_0}\right)^{2\, v_0^2}\exp{\left(-\frac{2 v_0^4}{a_M r}\right)}
\end{equation}
By identifying,
\begin{equation}
v_0^2=\sqrt{a_M M_B},
\end{equation}
\begin{figure}
    \centering
    \includegraphics[scale=0.5]{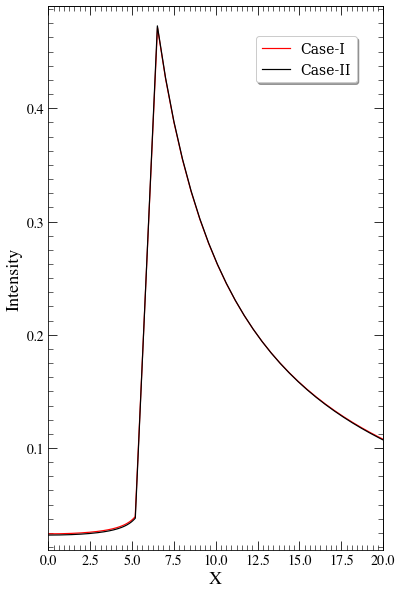}
    \caption{Plots of the corresponding intensities using the infalling gas as seen by a distant observer in black hole spacetimes surrounded by baryonic and apparent dark matter for the Case I and Case II, using $r_c \sim 10^6$, $M_B=10^3$ and $v_0=10^{-3}$, respectively.   }
\end{figure}
\begin{figure*}
    \includegraphics[scale=0.4]{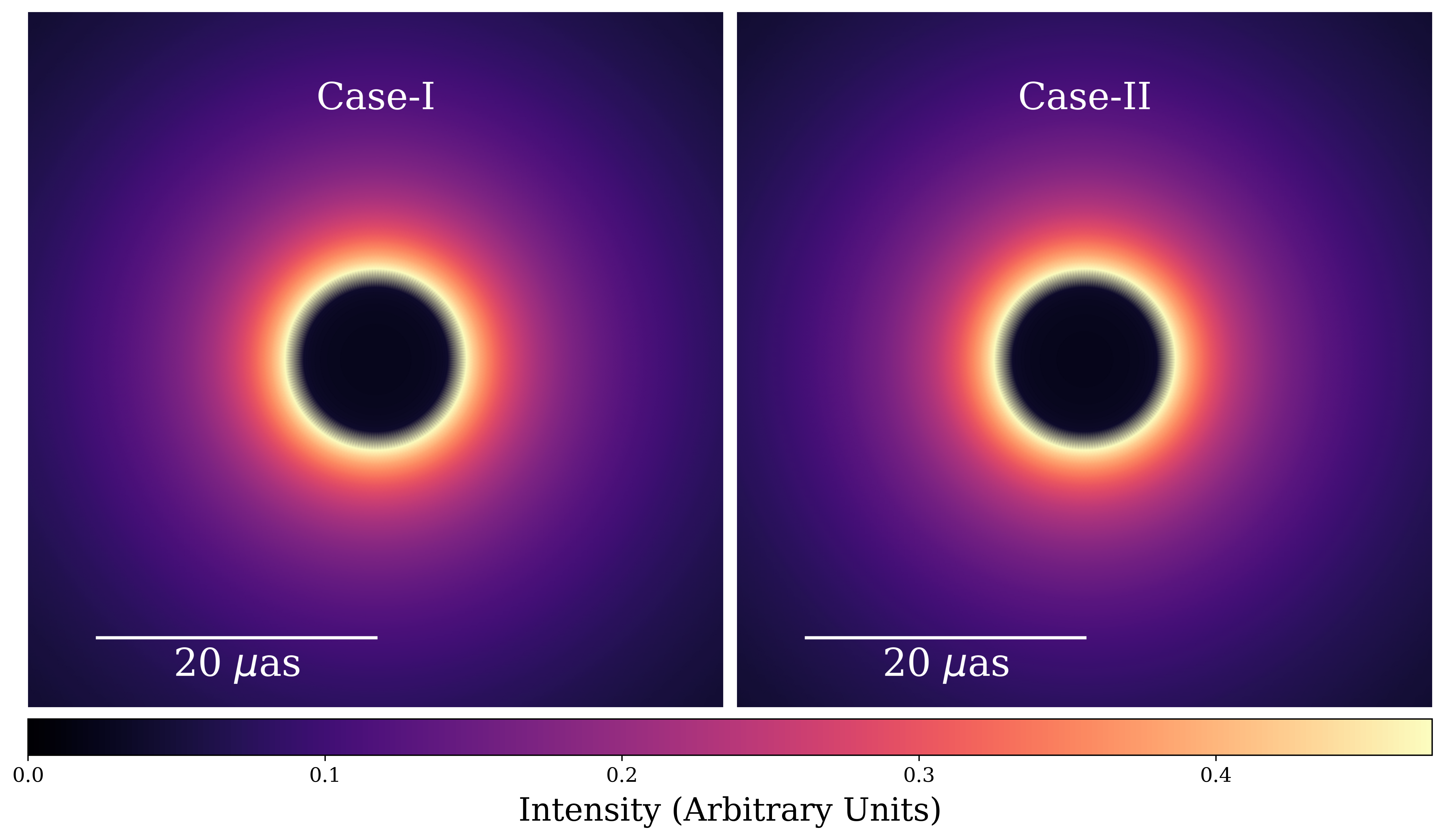}
    \caption{Images of shadows along with corresponding intensities using infalling gas as seen by a distant observer in a black hole spacetime for the Case I and Case II, using $r_c \sim 10^6$, $M_B=10^3$ and $v_0=10^{-3}$, respectively. }
\end{figure*}

We now consider black holes surrounded by apparent dark matter halo and baryonic matter. This space-time contribution can be obtained using corresponding energy-momentum tensors describing the total surrounding matter in Einstein field equation given by 
\begin{equation}
R^{\nu}_{~\mu}-\dfrac{1}{2}\delta^{\nu}_{~\mu}R=8 \pi T^{\nu}_{~\mu}.
\label{SPBH7}
\end{equation}
The space-time metric including black hole is thus given by 
\begin{equation}
ds^{2}=-\left(f(r)+F_{1}(r)\right)dt^{2}+\frac{dr^{2}}{g(r)+G_{1}(r)}+r^{2}d\Omega^2,
\label{SPBH9}
\end{equation}
where 
\begin{equation}
F(r) = f(r)+F_{1}(r),\quad G(r) = g(r)+G_{1}(r)
\end{equation}
where $d\Omega^2=d\theta^2+\sin^2\theta d\phi^2$. For any given dark matter density profile, we can obtain the corresponding space-time. In this way, it can be shown that $F_1(r)=G_1(r)=-2m/r$ (see, for details \cite{Xu:2018wow}). Thus, in the the general case having a black hole solution surrounded by matter with the assumption $f(r)=g(r)$, it can be shown that
\begin{equation}
F(r)=C\, r^{2 v_0^2}\,\left({1+\frac{r_c}{r}}\right)^{-\frac{2M}{r_c}}-\frac{2m}{r}.
\end{equation}
For very large but finite distance as we pointed out, we need to introduce a cut-off distance and by considering a Taylor series around $v_0$, we find
\begin{equation}
F(r)=1+2 v_0^2 \ln(r)+\hdots
\end{equation}
This result is consistent with what was argued in our Model I. However, there are few differences between these two models. Firstly, in our Model I, the black hole horizon was assumed  to be unchanged due to the surrounding matter. Secondly, in our Model II, the situation is different, namely, here the black hole horizon is affected by the surrounding matter. Thirdly, in the Model I we have $F(r) \neq G(r)$, which is not the case in our Model II.

\subsubsection{Case IV}
In our final case, let us consider the density profile for baryonic matter given by Eq. (2.11) along with the apparent dark matter profile given by Eq. (2.13) yielding
\begin{equation}
v^2_{tg}(r)=\frac{M_B r  }{(r+r_c)^2}+\frac{v_0^2 r^{3/2}}{(r+r_c)^{3/2}}\left(1+\frac{3 r_c}{2 r}  \right).
\end{equation}
Next, solving Eq. (2.26) we obtain
\begin{equation}
f(r)= C e^{-\frac{2 M_B}{r+r_c}}  \exp{\Big[ \frac{2 v_0^2 \left( \Delta_1 (r+r_c)+\sqrt{r(r+r_c)}\right)} {r+r_0}\Big]}
\end{equation}
where $\Delta_1$ is given by Eq. (2.18). Finally, assumption $f(r)=g(r)$, we  can add a black hole by following the same approach as in the Case III, yielding
\begin{eqnarray}\notag
F(r)&=& C e^{-\frac{2 M_B}{r+r_c}}  \exp{\Big[ \frac{2 v_0^2 \left(  \Delta_1 (r+r_c)+\sqrt{r(r+r_c)}\right)} {r+r_c}\Big]}\\
&-&\frac{2m}{r}.
\end{eqnarray}
Taking a Taylor series around $v_0$ we can check the consistency with Eq. (2.16), at least in leading order terms.  As we already pointed out, the main difference is that, in this model, the black hole event horizon  is affected by the surrounding matter. 

\section{Black hole shadow Radius}
Here, we are interested in investigating the shadow of black hole solution surrounded by matter. To do so, we start from Hamilton-Jacobi method for null geodesics in the black hole spacetime written as \cite{Perlick:2015vta}
\begin{equation}
\frac{\partial S}{\partial \sigma}+H=0,
\end{equation}
in which $S $ is the Jacobi action and  $\sigma $  is some affine parameter along
the geodesics.  If we consider a photon along null geodesics in our spherically symmetrical spacetime surrounded by matter, one can show that the Hamiltonian can be written as
\begin{equation}
\frac{1}{2}\left[-\frac{p_{t}^{2}}{F(r)}+G(r)p_{r}^{2}+\frac{p_{\phi}^{2}}{r^{2}}\right] =0.
\label{EqNHa}
\end{equation}

Due to the spacetime symmetries related to the coordinates $t$ and $\phi$, there are two constants of motion defined as follows
\begin{eqnarray}
p_{t}&\equiv\frac{\partial H}{\partial\dot{t}}=-E.\\
p_{\phi}&\equiv\frac{\partial H}{\partial\dot{\phi}}=L.
\end{eqnarray}

In the last two equations, $E$ and $L$ are the energy and the angular momentum of the photon, respectively. Next, the  circular and unstable orbits are related to the the maximum value of effective potential in terms of the following conditions
\begin{equation}
V_{\rm eff}(r) \big \vert_{r=r_{p}}=0,  \qquad \frac{\partial V_{\rm eff}(r)}{\partial r}%
\Big\vert_{r=r_{p}}=0,  
\end{equation}
 
Without going into details here, one can now show the following equation of motion 
\begin{equation}
\frac{dr}{d\phi}=\pm r\sqrt{G(r)\left[\frac{r^{2}f(R)}{R^{2}f(r)} -1\right] }. 
\end{equation}

Let us consider a light ray sent from a static observer located at a position $r_{0} $ and transmitted with an angle $\vartheta$ with respect to the radial direction. We, therefore, have \cite{Perlick:2015vta}
\begin{equation}
\cot \vartheta =\frac{\sqrt{g_{rr}}}{g_{\phi\phi}}\frac{dr}{d\phi}\Big\vert%
_{r=r_{0}}.  \label{Eqangle}
\end{equation}

Finally,  the relation for shadow radius of the black hole as observed by a static observer at the position $r_0$  can be shown as 
\begin{equation}
r_{s}=r_{0}\sin \vartheta =R\sqrt{\frac{f(r_{0})}{f(R)}}\Bigg\vert_{R=r_{p}}.
\end{equation}
where $r_p$ represents the photon sphere radius.  The apparent shape of a shadow as seen by the observer can be obtained by a stereographic projection in terms of celestial coordinates $X$ and $Y$ which are defined by
\begin{equation}\notag
X = \lim_{r_{0}\longrightarrow \infty} \left( -r_{0}^{2}\sin
\theta_{0}\frac{d\phi}{dr}\Big\vert_{(r_{0},\theta_{0})}\right), 
\end{equation}
\begin{equation}
Y = \lim_{r_{0}\longrightarrow \infty} \left( r_{0}^{2}\frac{d\theta}{%
dr}\Big\vert_{(r_{0},\theta_{0})}\right).
\end{equation}
It is worth noting that $( r_{0},\theta_{0}) $ are the position coordinates of the observer located at a far distance from the black hole. In the next section, we will consider a spherically symmetrical accretion model of infalling gas along with shadow images.  

\begin{figure}
    \centering
    \includegraphics[scale=0.5]{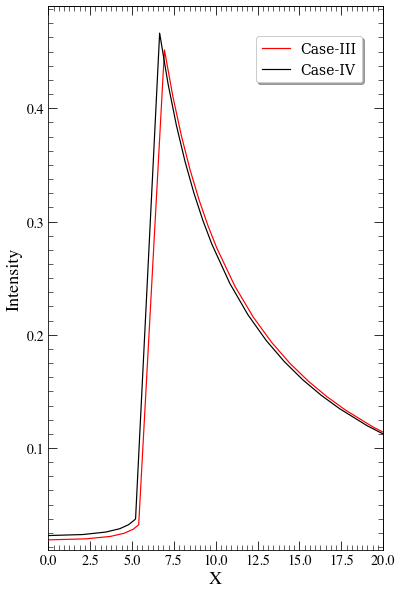}
    \caption{Plots of the corresponding intensities using the infalling gas as seen by a distant observer for the Case III and the Case IV, using $r_c \sim 10^6$, $M_B=10^3$ and $v_0=10^{-3}$, respectively. }
\end{figure}
\begin{figure*}
    \includegraphics[scale=0.4]{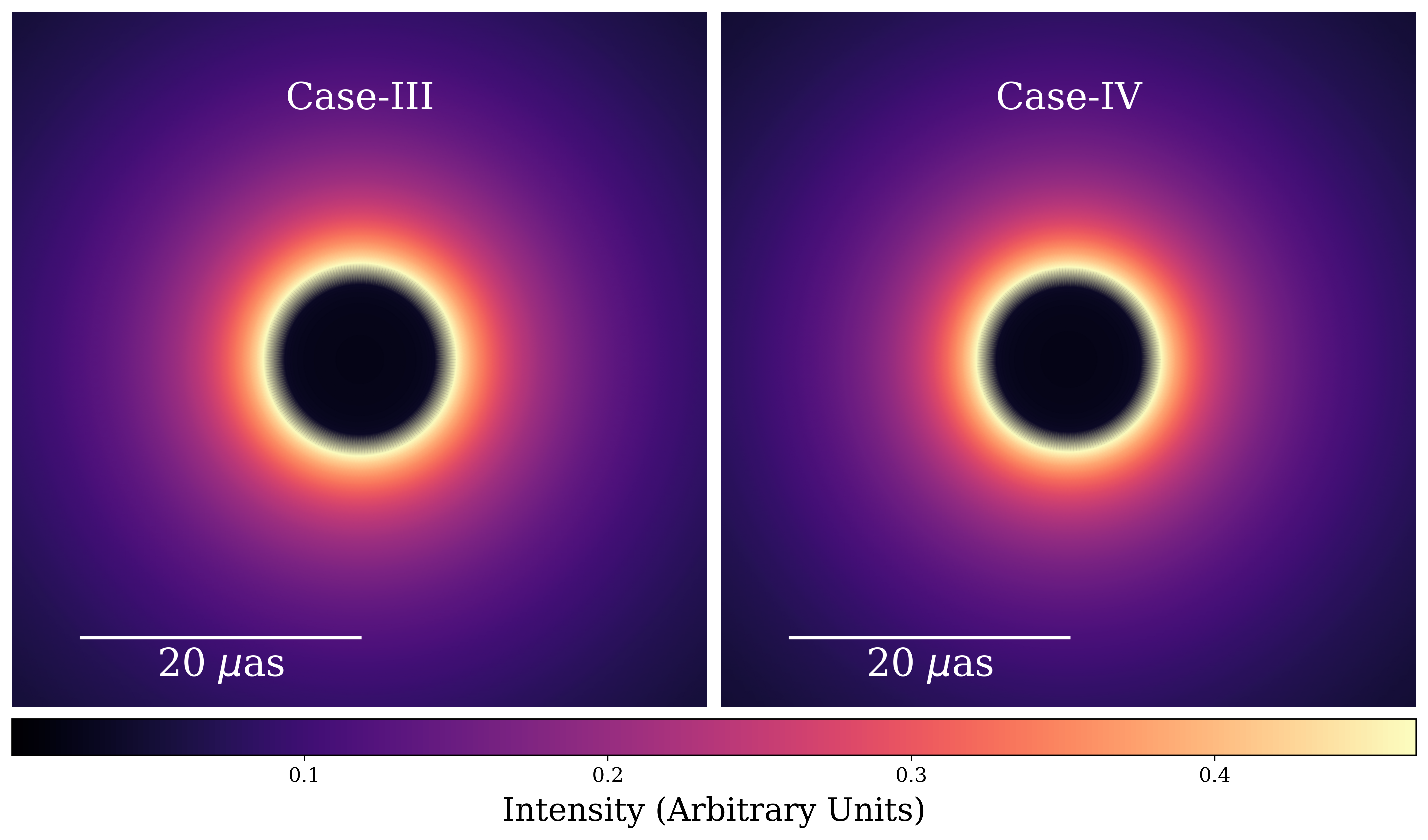}
    \caption{Images of shadows along with corresponding intensities using infalling gas as seen by a distant observer in a black hole spacetime using  Case III and Case IV, using $r_c \sim 10^6$, $M_B=10^3$ and $v_0=10^{-3}$, respectively.}
\end{figure*}

Using Eq. (3.8) we compute the corresponding shadow radius for the black hole  using $r_c \sim 10^6$, $M_B=10^3$ and $v_0=10^{-3}$. For the Case I and Case II we obtain $r_s=5.25$ and  $r_s=5.20$, respectively. For the Case III and Case IV we obtain for the shadow radius $r_s=5.39$ and  $r_s=5.21$, respectively. It should be noted that, since the spacetime geometry is not asymptotically flat, one must incorporate the finite distance corrections. In particular, this fact becomes significant and affects more the shadow radius for the Case I and Case II models. In the above examples we have used the M87* black hole case. Finally, we see that the above values are bigger compared to the Schwarzschild black hole with $r_s=5.196$, while in Fig. 2, we show the difference in the size of the shadow radius between those models. 

\section{Optically thin radiating and infalling accretion flow around the Black hole}
In this section, we consider a realistic and simple model of accretion flow surrounding the black hole; an optically thin, radiating accretion flow surrounding the object. The intensity of emitting region requires some assumption about the radiating processes and emission mechanisms. The observed specific intensity $I_{\nu 0}$ at the observed photon frequency $\nu_\text{obs}$ at the point $(X,Y)$ of the observer's image (usually measured in $\text{erg} \text{s}^{-1} \text{cm}^{-2} \text{str}^{-1} \text{Hz}^{-1}$) is given by \cite{PhysRevD.87.107501}
\begin{equation}
    I_{obs}(\nu_{obs},X,Y) = \int_{\gamma}\mathrm{g}^3 j(\nu_{e})dl_\text{prop},  
\end{equation}
where $\mathrm{g} = \nu_{obs}/\nu_{e}$ is the red-shift factor, $\nu_{e}$ is the photon frequency as measured in the rest-frame of the emitter, $j(\nu_{e})$ is the emissivity per unit volume in the rest-frame of the emitter, and $dl_\text{prop} = k_{\alpha}u^{\alpha}_{e}$ is the infinitesimal proper length as measured in the rest-frame of the emitter. The red-shift factor is evaluated from 
\begin{equation}
    \mathrm{g} = \frac{k_{\alpha}u^{\alpha}_{\text{obs}}}{k_{\beta}u^{\beta}_{e}}  
\end{equation}
where $k^{\mu}$ is the four-velocity of the photons, $u^{\alpha}_{e}$ four-velocity of the accreting gas emitting the radiation, $u^{\mu}_{\text{obs}}$ = $(1,0,0,0)$ and $\lambda$ is the affine parameter along the photon path $\gamma$. Here, $\gamma$ in the integral indicates that the integral has to be evaluated along the path of the photon (null geodesics).
Here, we are considering a simplistic case of accreting gas where it is in radial free fall .
For specific emissivity, we assume a simple model in which the emission is monochromatic with emitter's-rest frame frequency $\nu_{\star}$ and the following power law profile.
\begin{equation}
    j(\nu_{e}) \propto \frac{\delta(\nu_{e}-\nu_{\star})}{r^2},
\end{equation}
where $\delta$ is the Dirac delta function. 
Integrating the intensity over all the observed frequencies, we obtain the observed flux
\begin{equation}
    F_{obs}(X,Y) \propto -\int_{\gamma} \frac{\mathrm{g}^3 k_t}{r^2k^r}dr.
\end{equation}

Firstly, we explore the shadow images for the Case I and Case II. In Fig. 3 and Fig. 4 we show the intensities and the shadow images using the infalling gas model as seen by a distant observer in a  black hole spacetime surrounded by matter and apparent dark matter. It can be seen that in both cases, the effect of the surrounding matter on the shadow radius is very small and increases with the increase in the surrounding mass. As we have already noted, one must use finite distance corrections to compute the shadow radius since the spacetime is not asymptotically flat.  On the other hand, in Fig. 5 and Fig. 6,  we show the intensities and shadow images for the black hole described by  the Case III and Case IV, respectively. Here we observe that the effect of the surrounding matter on the shadow radius is bigger compared to the Model I, while the intensity is smaller. This difference is a direct result of the fact that the black hole event horizon is affected by the surrounding matter in the Model II.  For Model I, we find that the effects becomes significant once the surrounding mass is of the order $M_B \sim 10^4-10^5$. Model II is more sensitive to the shadow radius and the effects becomes significant once the surrounding mass is $M_B \sim 10^3-10^4$. This shows that the shadow images can be used as an indirect tool to detect matter around black holes.

\begin{figure*}
    \includegraphics[scale=0.62]{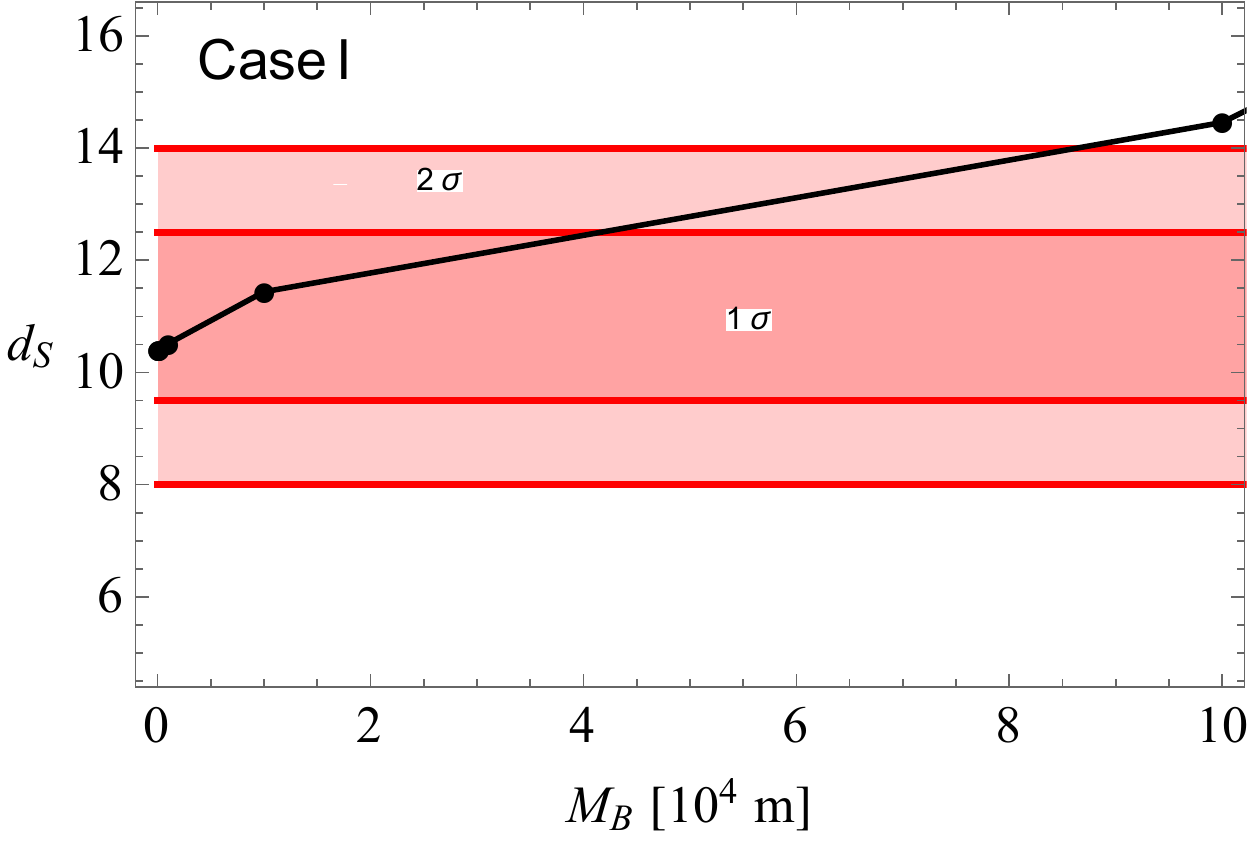}
     \includegraphics[scale=0.62]{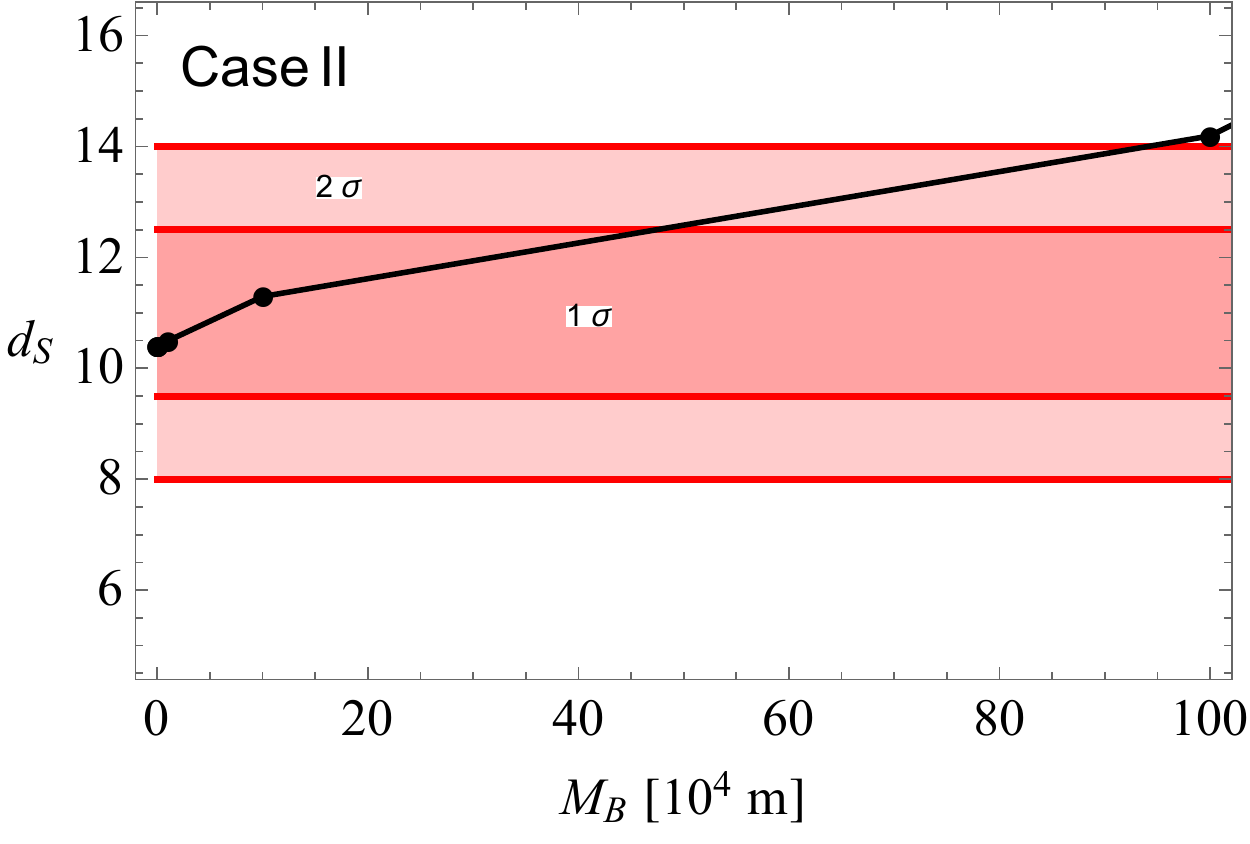}
    \caption{The regions of parameter space of the diameter of the black hole shadow for the Case I and Case II  within $1\sigma$ and $2\sigma$ uncertainties, respectively. Note that $M_B$ is measured in units of black hole mass with $m=1$. }
\end{figure*}
\begin{figure*}
    \includegraphics[scale=0.62]{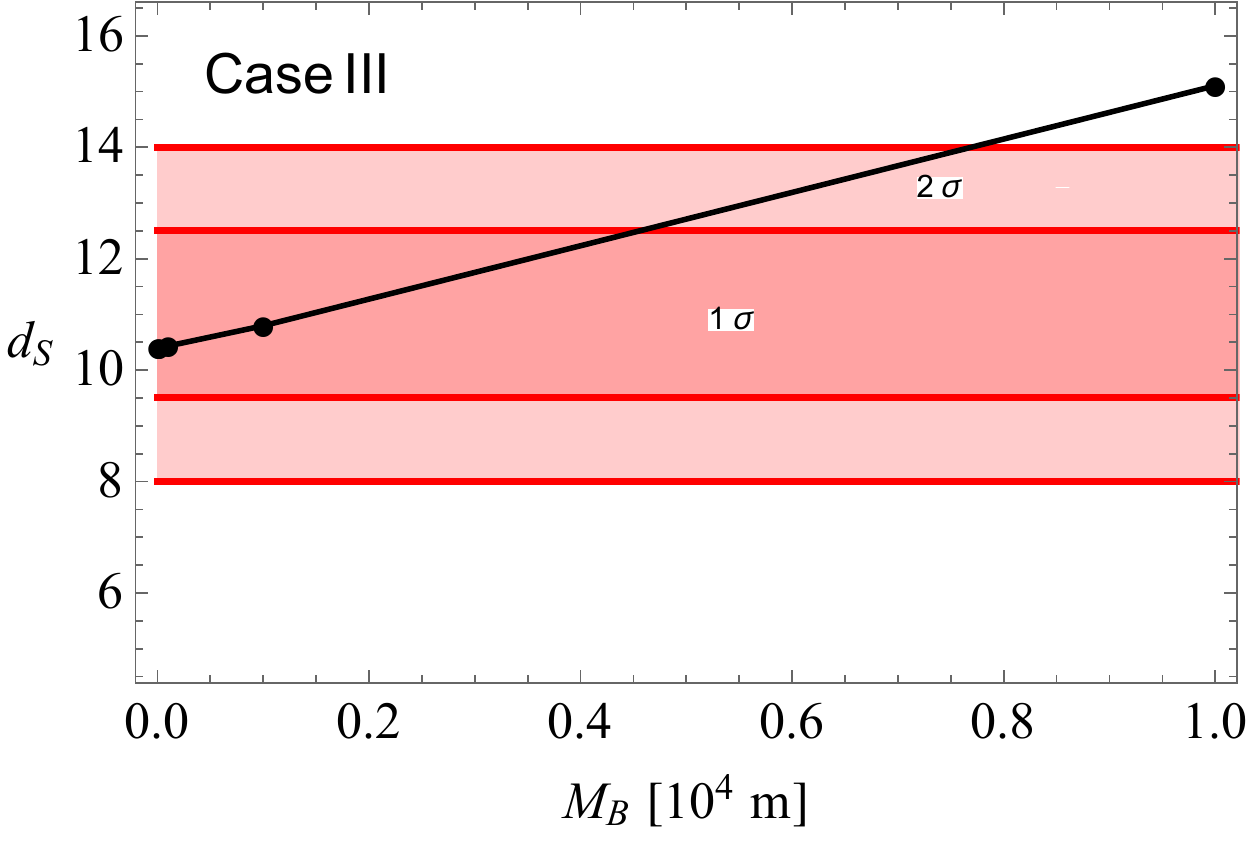}
     \includegraphics[scale=0.62]{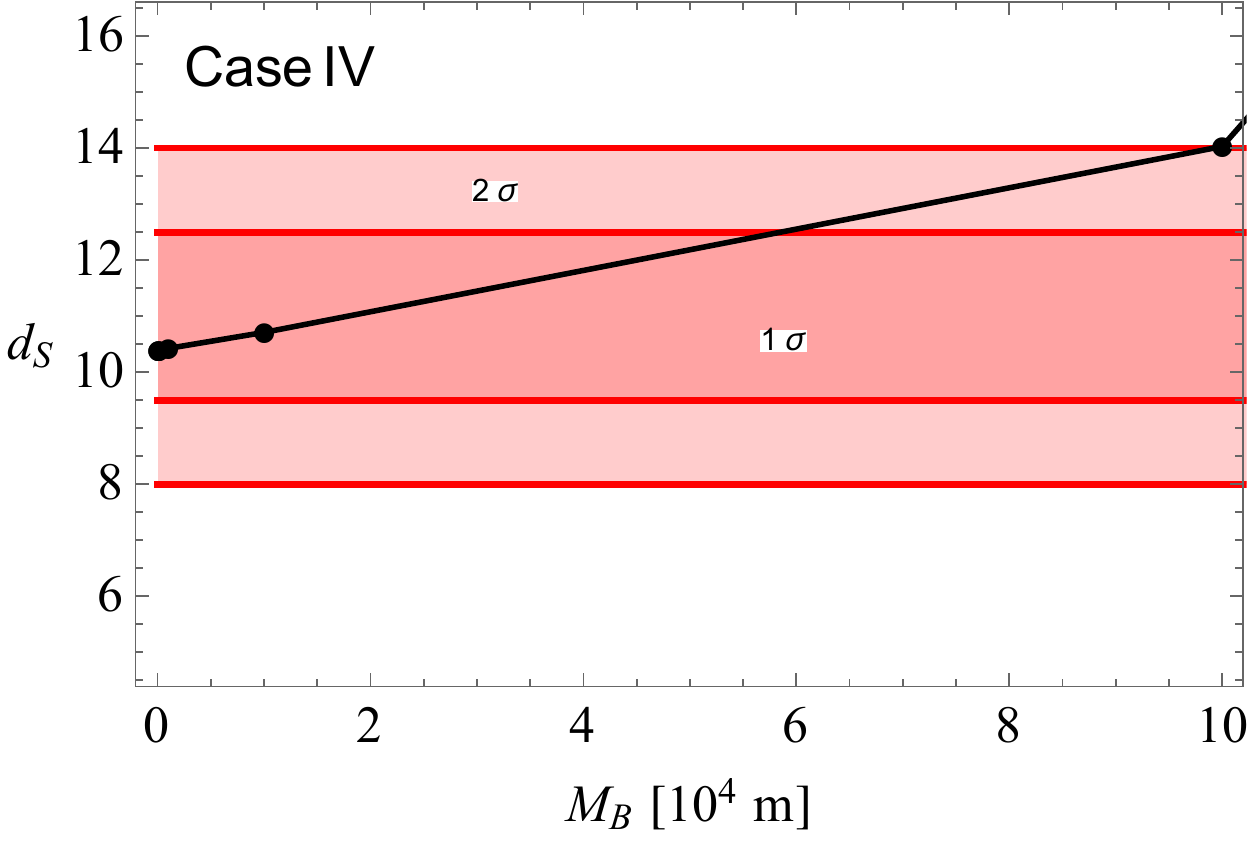}
    \caption{The regions of parameter space of the diameter of the black hole shadow for the Case III and Case IV within $1\sigma$ and $2\sigma$ uncertainties, respectively. Note that $M_B$ is measured in units of black hole mass with $m=1$. }
\end{figure*}

\section{Observational constraints}
 In this section, we shall use the  reported angular size of the black hole shadow in the M87 galactic center reported by  EHT  $\theta_s = (42 \pm 3)\mu as$, along with the distance to M87 given by  $D = 16.8 $ Mpc, and the mass of M87 central object $M = 6.5 \times 10^9$ M\textsubscript{\(\odot\)} to constrain the baryonic mass around the black hole M87. In order to constrain the surrounding baryonic mass $M_B$, for simplicity,  we are going to neglect the rotation.  Next, the diameter of the shadow in units of mass $d_{M87}$ is given by \cite{Allahyari:2019jqz}
\begin{eqnarray}
d_{M87}=\frac{D \,\theta_s}{M_{87}}=11.0 \pm 1.5.
\end{eqnarray}
We are further going to study the following separate cases:
\subsection{Model I}
In this model we have two specific cases; Case I described by mass profile (2.7), and Case II described by the mass profile (2.13). Within $1\sigma $ confidence, we have the interval $9.5 \leq d_{M87} \leq 12.5$, whereas within $2\sigma $ uncertainties, we have $8 \leq d_{M8} \leq 14 $. In Fig. 7, we show the regions of parameter space of the diameter of the shadow for the Case I and Case II, respectively.  Within $2\sigma$ confidence, we find the upper bound for the surrounding  baryonic mass to be  $  M_B \leq 8.6 \times 10^4$.  On the other hand, for Case II, within $2\sigma$ confidence, we find the upper bound for the surrounding  baryonic mass to be  $  M_B \leq 95 \times 10^4$ measured in units of the black hole mass.  

\subsection{Model II}
Finally, let us consider Model II, and see the effect of baryonic mass on the shadow. In Fig. 8, we show the regions of parameter space of diameter of the shadow for the Model II.  For the Case III, within $2\sigma$ confidence, we find the upper bound for the surrounding  baryonic mass to be  $  M_B \leq 7.6 \times 10^3$.  For the Case IV, within $2\sigma$ confidence, we find the upper bound for the surrounding  baryonic mass to be  $  M_B \leq 10 \times 10^4$. From all these plots, we conclude that the effect becomes significant when the surrounding baryonic matter is comparable $M_B \sim  10^4$. We should point out that our analysis is based on the assumption that most of the baryonic matter is located near the galactic center. Note that the baryonic mass $M_B$ in all plots is measured in units of black hole mass defined as $m=1$.

\section{Conclusions}
In this paper, we studied the shadow images of black holes in Verlinde’s Emergent Gravity. Toward this purpose, we considered a black hole surrounded by baryonic mass and an optically-thin gas medium in radial free fall. In order to study the influence of surrounding matter on the photon sphere, we assumed that most of the baryonic matter in the galaxy is located near the galactic center. In particular, we used two different models to construct the spacetime metric near the black hole. In the first toy model, we considered the simplest case, namely a power law mass profile for the baryonic mass, then we extended our analysis by assuming a different power law mass profile known as  the Hernquist model.  In both cases, we studied not only the effect of baryonic mass but also the effect of apparent dark matter on the shadow and the corresponding intensities. We have shown that, the surrounding matter increases the shadow radius while the effect becomes significant when the surrounding baryonic matter is $M_B\sim 10^3-10^5$ in units of black hole mass. It is also shown that  intensity of the electromagnetic flux radiation observed by distant observer decreases with the increase in mass. This of course can be explained by the fact that as the black hole spacetime gets distorted by the extra effect coming from the baryonic and apparent dark matter, the number of photons captured by the black hole increases. As a result, we end up with a smaller value for the intensity at a large distance.  In addition to that, the constant $r_c$ in the power law model, is assumed to be in the range of an observational value, that is of the order of kpc. Moreover we showed  the influence of baryonic/apparent dark matter on the electromagnetic radiation emitted from spherical accretion medium which was assumed to be an optically-thin region surrounding the black hole. 
In the present work, in Model I we used TOV equation to construct our black hole metric surrounded by matter. In Model II, we used the tangential velocity of the test particle to construct the spacetime metric having a black hole at the center. Similarly, we found that increasing the surrounding baryonic mass increases the shadow radius. As a special case of the power law we obtained the effect of constant mass function in Verlinde’s theory and argued that the corresponding apparent dark matter to be similar to the isothermal dark matter profile in ordinary general relativity. In that sense, the shadow images of a black hole surrounded by a constant baryonic matter and apparent dark matter in Verlinde's theory, are almost indistinguishable from shadow images obtained in ordinary general relativity having dark matter described as an isothermal sphere. \\
Note that in a more realistic situations, black holes are expected to rotate and it is well known that the angular momentum affects the apparent shape of the black hole shadow. In our case, we have a black hole surrounded by matter/dark matter and, unfortunately,  there is no general method to find an exact solutions with a rotating black holes surrounded by matter. However, in such situations, one can only find an effective rotating metric by applying a complex coordinate transformation, or the Newman–Janis algorithm. We plan to expand this study in more details in the near future by introducing a rotation in our spacetime and also by considering different radii for the inner edge of the surrounding matter. On the other hand, since we analyzed the M87* black hole in the present work, it is quite remarkable that the detected diameter of M87* shadow is quite consistent with that of the Schwarzschild black hole case. In particular, this can be linked with the viewing angle $\theta_0=17^{0}$, such an angle is supported by the so-called Blandford--Znajek mechanism (see, \cite{Allahyari:2019jqz}). For such a viewing angle the effect of rotation is small therefore the shadow images of the M87* black hole are in good agreement with the shadow images given in Fig. 2.

\section*{Acknowledgements}
We would like to thank members of International Centre for Cosmology (ICC), Charusat University for their valuable and helpful discussions related to ray-tracing algorithm. The authors also thank the referee for valuable comments and suggestions. In addition, Saurabh would like to thank Saher and Priyamvada for the motivation and support required during the work.

%\label{lastpage}
\end{document}